\documentclass[prd,twocolumn,showpacs,preprintnumbers,amsmath,amssymb]{revtex4}

\newcommand{\br}{{\bf r}}
\newcommand{\bu}{{\bf u}}
\newcommand{\bw}{{\bf w}}

\begin{document}

\title{Maxwell's Multipole Vectors and the CMB}

\author{Jeff Weeks}
\affiliation{15 Farmer Street, Canton NY 13617 USA}%
\email{www.geometrygames.org/contact.html}

\begin{abstract}
    \noindent The recently re-discovered multipole vector approach
    to understanding the harmonic decomposition of the
    cosmic microwave background traces its roots
    to Maxwell's {\it Treatise on Electricity and Magnetism}.
    Taking Maxwell's directional derivative approach
    as a starting point, the present article develops
    a fast algorithm for computing multipole vectors,
    with an exposition that is both simpler and better
    motivated than in the author's previous work.

    Tests show the resulting algorithm, coded up as
    a Mathematica notebook, to be both fast and robust.
    This article serves to announce the free availability
    of the code.

    The algorithm is then applied to the much discussed
    anomalies in the low-$\ell$ CMB modes, with sobering results.
    Simulations show the quadrupole-octopole alignment to be unusual
    at the $98.7\%$ level, corroborating earlier estimates
    of a 1-in-60 alignment while showing recent reports
    of 1-in-1000 (using non-unit-length normal vectors)
    to have unintentionally relied on the near orthogonality
    of the quadrupole vectors in one particular data set.
    The alignment of the quadrupole and octopole vectors
    with the ecliptic plane is confirmed at better than
    the $2\sigma$ level.

\end{abstract}


\maketitle

\section{Introduction}

The Wilkinson Microwave Anisotropy Probe (WMAP) provides an
unprecedented view of the Cosmic Microwave Background (CMB) sky.
WMAP's first-year observations~\cite{WMAP-Bennett}, while
generally matching researchers' expectations in the high-$\ell$
portion of the CMB's spherical harmonic decomposition, yield
suspicious results in the low-$\ell$ portion of the spectrum.
Almost immediately after the first-year data release, Tegmark and
colleagues noticed that the reported quadrupole ($\ell$=2) and
octopole ($\ell$=3) align~\cite{Tegmark}, and Eriksen~et~al. found
statistical inconsistencies between the northern and southern
ecliptic hemispheres~\cite{Eriksen1}.  Eriksen and colleagues
later found that while the $\ell=4$ component is generic, the
$\ell=5$ component is unusually spherical at the $3\sigma$ level
while the $\ell=6$ component is unusually planar at the $2\sigma$
level~\cite{Eriksen2}.

To make sense of these anomalies in a systematic way, Copi~et~al.
introduced ``a novel representation of cosmic microwave anisotropy
maps, where each multipole order $\ell$ is represented by $\ell$
unit vectors pointing in directions on the sky and an overall
magnitude''~\cite{Copi}.  The beauty of this scheme is that the
CMB alone determines the multipole vectors, without reference to
the coordinate system.  This contrasts sharply with the more
common $a_{\ell m}$ representation of spherical harmonics, which
depends strongly on the coordinate system.  The coordinate-free,
geometrical nature of the multipole vectors makes it trivially
easy to test for alignments between modes (for example the
quadrupole and octopole align with each other) and with external
reference points (the quadrupole and octopole align with the CMB
dipole)~\cite{Schwarz}.

Inspired by these results, Katz and Weeks went on to recast the
traceless tensor methods of Refs.~\cite{Copi,Schwarz} into the
language of homogeneous harmonic polynomials, and within that
context to prove the existence and uniqueness of the multipole
vectors and to devise an algorithm for computing
them~\cite{KatzWeeks}.

Even though Copi~et~al. fancied themselves the discovers of
multipole vectors~\cite{MPVectors} and Katz-Weeks fancied
themselves to be breaking new ground in the polynomial approach,
the truth is much older.  The true discover of multipole vectors
was James Clerk Maxwell in his 1873 {\it Treatise on Electricity
and Magnetism}~\cite{Maxwell}~!  Maxwell's starting point was, of
course, electromagnetism.  He began with the $\frac{1}{r}$
potential surrounding a point charge (a {\it monopole}) and asked
what happens when two monopoles of opposite sign get pushed
together. Elementary reasoning shows the potential of the
resulting {\it dipole} to be given (up to rescaling) by the
directional derivative $\nabla_{\bu_1}\frac{1}{r}$, where $\bu_1$
is the direction along which the two monopoles approach one
another. Similarly, pushing together two such dipoles with
opposite signs gives (up to rescaling) a {\it quadrupole} with
potential $\nabla_{\bu_2}\nabla_{\bu_1}\frac{1}{r}$, where $\bu_2$
is the direction along which the dipoles approach, and so on.
Maxwell recognized these potentials as solutions of Laplace's
equation, i.e. spherical harmonics.  Because his expression
$\lambda\nabla_{\bu_\ell}\cdots\nabla_{\bu_1}\frac{1}{r}$ contains
$2\ell+1$ degrees of freedom, and because a simple dimension
counting argument indicates that a generic harmonic polynomial of
degree~$\ell$ also has $2\ell+1$ degrees of freedom, Maxwell
concluded -- with less than perfect rigor but nevertheless
correctly -- that his method yields all spherical harmonics. \\

Section~\ref{SubsectionEquivalence} establishes, by purely
elementary means, the equivalence between Maxwell's multipole
vectors and the vectors used in more recent
work~\cite{Copi,Schwarz,KatzWeeks}.
Section~\ref{SubsectionAlgorithm} sheds fresh light on the fast
algorithm for computing multipole vectors, providing a simple,
natural motivation for a construction that felt unnatural and
ad~hoc in Ref.~\cite{KatzWeeks}.
Section~\ref{SectionImplementation} announces the release of a
Mathematica package implementing the fast algorithm, and analyzes
its efficiency and stability.  Finally,
Section~\ref{SectionApplications} applies the fast algorithm to
the first-year WMAP data, finding both the quadrupole-octopole
alignment and the alignments with the ecliptic plane to be
significant only at the $99\%$ level, not at the higher levels
claimed in Refs.~\cite{Schwarz,KatzWeeks}.

\section{Theory}
\label{SectionTheory}

\subsection{Proof of equivalence}
\label{SubsectionEquivalence}

While Maxwell's directional derivatives provide a welcome breath
of fresh air for researchers toiling with more difficult
approaches, namely traceless tensors~\cite{Copi} and factored
polynomials~\cite{KatzWeeks}, one must nevertheless prove that
Maxwell's multipole vectors are indeed the same vectors that
appear in those other approaches.  Mark Dennis has successfully
applied Fourier methods~\cite[Appendix~A]{Dennis1} to prove the
equivalence of the polynomial and Maxwell interpretations. Here we
obtain the same result by more elementary means, keeping to the
spirit of Maxwell's original work~\cite[Chap.~IX]{Maxwell}.

Maxwell expresses a spherical harmonic as
\begin{equation}
    f_\ell(x,y,z)
    = \nabla_{\bu_\ell} \cdots \nabla_{\bu_2} \nabla_{\bu_1} \;
    \frac{1}{r},
    \label{Maxwell}
\end{equation}
where $r = \sqrt{x^2 + y^2 + z^2}$.  Observe the simple pattern
that emerges as we apply the directional derivatives one at a
time,
\begin{eqnarray}
    f_0 &=& \frac{1}{r}\nonumber\\
    f_1 &=& \nabla_{\bu_1}f_0 = \frac{(-1)(\bu_1 \cdot \br)}{r^3}\nonumber\\
    f_2 &=& \nabla_{\bu_2}f_1 = \frac{(3\cdot 1) (\bu_1 \cdot \br)(\bu_2 \cdot \br) + r^2(-\bu_1 \cdot \bu_2)}{r^5}\nonumber\\
    f_3 &=& \nabla_{\bu_3}f_2 = \frac{(-5\cdot 3\cdot 1) (\bu_1 \cdot \br)(\bu_2 \cdot \br)(\bu_3 \cdot \br) + r^2(...)}{r^7}\nonumber\\
\end{eqnarray}
where boldface $\br = (x,y,z)$ while plain $r =
\sqrt{\br\cdot\br}$ as before.  The ellipsis (\dots) marks a
polynomial whose form does not interest us.

Let $P_\ell$ denote the polynomial in the numerator of each
$f_\ell$.  The action of $\nabla_{\bu_{\ell}}$ may be written
explicitly as
\begin{eqnarray}
    f_{\ell}
    &=& \nabla_{\bu_{\ell}}f_{\ell - 1}
    = \nabla_{\bu_{\ell}}\left( \frac{P_{\ell-1}}{r^{2\ell - 1}}\right)\nonumber\\
    &=& \frac{-(2\ell - 1)(\bu_\ell \cdot \br) P_{\ell-1}
        + r^2 (\bu_{\ell}\cdot\nabla P_{\ell-1})}
        {r^{2\ell + 1}}.
\label{QuotientRuleEffect}
\end{eqnarray}
It is now obvious at a glance that
formula~(\ref{QuotientRuleEffect}) takes
\begin{equation}
    f_{\ell - 1} = \frac{(-1)^{\ell-1}\;(2\ell - 3)!!\;(\bu_1 \cdot \br)\cdots(\bu_{\ell - 1} \cdot \br) + r^2(\dots)}{r^{2\ell - 1}}
\end{equation}
to
\begin{equation}
    f_{\ell} = \frac{(-1)^{\ell}\;(2\ell - 1)!!\;(\bu_1 \cdot \br)\cdots(\bu_{\ell} \cdot \br) + r^2(\dots)}{r^{2\ell + 1}}
    \label{FormulaFell}
\end{equation}
thus establishing by induction the validity of the latter for all
$\ell$.

We would now like to understand more deeply the relationship
between the rational function $f_\ell$ and the polynomial $P_\ell$
appearing in its numerator.  Following the method of electrical
inversion, which Maxwell~\cite[art.~162]{Maxwell} credits to
Thomson (Lord Kelvin) and Tait~\cite{Kelvin1} but
Axler~\cite{Axler} traces back to Kelvin's work two decades
earlier~\cite{Kelvin2}, define the {\it Kelvin transform}
$\tilde{f}$ of a function $f$ to be
\begin{equation}
    \tilde{f}(x,y,z)
    = \frac{1}{r} f(\frac{x}{r^2}, \frac{y}{r^2}, \frac{z}{r^2}).
\end{equation}
The Kelvin transform in effect reflects a function across the unit
sphere and then adjusts its amplitude to keep it harmonic
(to be proved below). \\

\noindent {\it Lemma 1.}  The functions $f$ and $\tilde f$ agree
on the unit sphere $r = 1$. \\

\noindent {\it Lemma 2.}  The Kelvin transformation $\tilde{\;}$
is an involution, that is,
\begin{eqnarray}
    f(x,y,z)
    \;&\tilde{\rightarrow}&\;
    \frac{1}{r} f(\frac{x}{r^2}, \frac{y}{r^2},\frac{z}{r^2})\nonumber\\
    \;&\tilde{\rightarrow}&\;
    \frac{1}{r}\frac{1}{\frac{r}{r^2}} f(
        \frac{\frac{x}{r^2}}{\frac{r^2}{r^4}},
        \frac{\frac{y}{r^2}}{\frac{r^2}{r^4}},
        \frac{\frac{z}{r^2}}{\frac{r^2}{r^4}})
    =
    f(x,y,z).
\end{eqnarray}
\\

\noindent {\it Lemma 3.}  If $f$ is harmonic, then so is $\tilde
f$.  Maxwell calls this the theorem of electrical
inversion~\cite[art.~129]{Maxwell}.  It is easily proved by a
direct calculation
\begin{equation}
    \nabla^2\left( \frac{1}{r} f(\frac{x}{r^2}, \frac{y}{r^2}, \frac{z}{r^2}) \right)
    = \frac{1}{r^5}(\nabla^2 f)(\frac{x}{r^2}, \frac{y}{r^2}, \frac{z}{r^2})
    = 0.
\end{equation}
The converse, that $\tilde f$ harmonic implies $f$ harmonic, is automatically true by Lemma 2. \\

Each multipole function $f_\ell = \frac{P_\ell}{r^{2\ell + 1}}$
has a polynomial numerator $P_\ell$ of homogeneous degree $\ell$,
so its Kelvin transform is easy to compute,
\begin{equation}
    \tilde f_\ell
    = \frac{1}{r} \frac{\frac{P_\ell}{r^{2\ell}}}{\frac{r^{2\ell + 1}}{r^{4\ell + 2}}}
    = P_\ell.
\end{equation}
In other words, the Kelvin transformation of $f_\ell$ extracts the
numerator $P_\ell$. Lemma 3 now implies that $P_\ell$ is a
harmonic polynomial in its own right.

Equation~(\ref{FormulaFell}) gives $P_\ell$ explicitly as
\begin{equation}
    P_{\ell} = (-1)^{\ell}\;(2\ell - 1)!!\;(\bu_1 \cdot \br)\cdots(\bu_{\ell} \cdot \br) + r^2(\dots).
    \label{Pell}
\end{equation}
Happily this is exactly the ``factored form'' whose existence and
uniqueness were established in Ref.~\cite{KatzWeeks} and
elsewhere.
Thus we have proven\\

\noindent {\it Proposition 4.}  Each harmonic function
\begin{equation}
    \nabla_{\bu_\ell} \cdots \nabla_{\bu_2} \nabla_{\bu_1} \;
    \frac{1}{r}
    \label{MaxwellsForm}
\end{equation}
defined by Maxwell agrees on the unit sphere with the ``factored
form''
\begin{equation}
    \lambda (\bu_1 \cdot \br) \cdots (\bu_{\ell} \cdot \br) + r^2 Q
    \label{FactoredForm}
\end{equation}
of Ref.~\cite{KatzWeeks}.  In particular, a given function employs
the same set of multipole vectors $\{\bu_1,
\dots, \bu_\ell\}$ in both representations. \\

The only remaining question is whether the mapping from Maxwell's
form~(\ref{MaxwellsForm}) to the factored
form~(\ref{FactoredForm}) is surjective.  For each choice of
$\{\bu_1, \dots, \bu_\ell\}$ the mapping hits exactly one factored
polynomial
\begin{equation}
    \lambda (\bu_1 \cdot \br) \cdots (\bu_{\ell} \cdot \br) + r^2 Q
    \label{Q}
\end{equation}
while infinitely many other polynomials
\begin{equation}
    \lambda (\bu_1 \cdot \br) \cdots (\bu_{\ell} \cdot \br) + r^2 Q'
    \label{Qprime}
\end{equation}
with $Q' \neq Q$, will not get hit.  So strictly speaking the
mapping is far from surjective.  The reason is that the factored
form applies to all homogeneous polynomials, while Maxwell's
method generates only homogeneous {\it harmonic} polynomials.  If
we restrict the factored form from all homogeneous polynomials to
only the harmonic ones, then the mapping becomes surjective for
the following reason.  The polynomials (\ref{Q}) and
(\ref{Qprime}) cannot both be harmonic, because then their
difference $r^2(Q - Q')$ would be harmonic as well, contradicting
the fact that no nonzero multiple of $r^2$ is harmonic (Endnote
\footnote{To prove that no nonzero multiple of $r^2 = x^2 + y^2 +
z^2$ is harmonic, consider a polynomial $(r^2)^n\, S$, where $n
\geq 1$ and $S$ contains no further factors of $r^2$.  If
$(r^2)^n\, S$ were harmonic, then computing $\nabla^2((r^2)^n\,
S)$ and setting it equal to zero would lead to $r^2 \nabla^2 S +
(4 n \,\rm{deg}(S) + 2n(2n+1))\,S = 0$, which would imply that $S$
contained another factor of $r^2$, contrary to assumption.
Therefore $(r^2)^n\, S$ cannot be harmonic.}).
Equation~(\ref{Pell}) gives a value of $Q$ for which (\ref{Q}) is
indeed harmonic, thus proving that the mapping from Maxwell's form
(\ref{MaxwellsForm}) to the factored form (\ref{FactoredForm}) has
as its image exactly the set of harmonic polynomials.

\subsection{The algorithm}
\label{SubsectionAlgorithm}

The most efficient algorithm for computing a spherical harmonic's
multipole vectors expresses a given spherical harmonic as a
homogeneous harmonic polynomial $P(x,y,z)$ and then looks at the
intersection of the graphs $P(x,y,z) = 0$ and $x^2 + y^2 + z^2 =
0$ in the complex projective plane.  The Katz-Weeks
paper~\cite{KatzWeeks} simply pulls this approach out of thin air,
because at the time my coauthor and I did not understand the
origins of our work.  It turns out that the key ideas in ``our''
algorithm first appeared in 1876 in J.J.~Sylvester's ``Note on
Spherical Harmonics''~\cite{Sylvester}. While Sylvester makes his
harsh feelings for Maxwell abundantly clear, he leaves his
algorithm and the motivation behind it decidedly unclear. Inspired
by Sylvester's paper, the following paragraphs lay out the
algorithm in elementary terms, speculating on what Sylvester's
reasoning might have been.

Sylvester's starting point was Maxwell's directional derivative
formulation~(\ref{Maxwell}) of a spherical harmonic, which reduces
to a polynomial~(\ref{Pell}), which for present purposes may be
written as
\begin{equation}
    P(x,y,z) = \lambda \prod_{i=1}^{\ell}(\bu_i \cdot (x,y,z))
        + (x^2 + y^2 + z^2)\cdot Q.
    \label{succinct}
\end{equation}
Let us take stock of the computational task that lies before us.
We are given the harmonic polynomial $P(x,y,z)$ of homogeneous
degree $\ell$, and we are asked to compute the linear factors
$\bu_i \cdot (x,y,z)$, but with no prior knowledge of the
remainder term $Q$.

If we could find a common solution to $P(x,y,z) = 0$ and $x^2 +
y^2 + z^2 =~0$, then such a solution would also satisfy at least
one of the linear equations $\bu_i\cdot (x,y,z) = 0$.  Of course
$x^2 + y^2 + z^2 = 0$ has no nontrivial solutions over the real
numbers, which motivates us to consider complex solutions instead.
Postponing the computational details, assume for the moment that
we have found a common solution $(\hat x, \hat y, \hat z)$ to
$P(x,y,z) = 0$ and $x^2 + y^2 + z^2 = 0$, which is therefore a
solution to $\bu_i\cdot (x,y,z) = 0$ for some $i$. Moreover,
because $\bu_i$ is real, the real and imaginary parts of $(\hat x,
\hat y, \hat z)$ individually satisfy the condition
\begin{eqnarray}
  \bu_i\cdot ({\rm Re}\,\hat x,\;{\rm Re}\,\hat y,\;{\rm Re}\,\hat z) = 0\phantom{.} \nonumber\\
  \bu_i\cdot ({\rm Im}\,\hat x,\;{\rm Im}\,\hat y,\;{\rm Im}\,\hat z) = 0.
\end{eqnarray}
Furthermore, the real vectors $({\rm Re}\,\hat x,\,{\rm Re}\,\hat
y,\,{\rm Re}\,\hat z)$ and $({\rm Im}\,\hat x,\,{\rm Im}\,\hat
y,\,{\rm Im}\,\hat z)$ must be linearly independent, because
otherwise $(\hat x, \hat y, \hat z)$ could not be a nontrivial
solution of $x^2 + y^2 + z^2 = 0$. Therefore their cross product,
when normalized to unit length, gives the desired multipole vector
\begin{eqnarray}
  \bu_i \sim ({\rm Re}\,\hat x,\;{\rm Re}\,\hat y,\;{\rm Re}\,\hat z)
     \times ({\rm Im}\,\hat x,\;{\rm Im}\,\hat y,\;{\rm Im}\,\hat z).
  \label{CrossProduct}
\end{eqnarray}
If we can find enough distinct solutions $(\hat x, \hat y, \hat
z)$, we will get all the multipole vectors $\bu_i$.

So how may one find the common solutions $(\hat x, \hat y, \hat
z)$ to $P(x,y,z) = 0$ and $x^2 + y^2 + z^2 = 0$ ?  Sylvester
applied his own {\it resolving equation} method, and Mathematica
might be employing a similar algorithm in its proprietary {\tt
NSolve} function.   Unfortunately {\tt NSolve} is slow.  If we
give Mathematica a little help we can speed the computation up
enormously.  Following Ref.~\cite{KatzWeeks}, we parameterize the
graph of $x^2 + y^2 + z^2 = 0$, which topologically is a Riemann
sphere $\mathbb{C}\mathrm P^1 \approx \mathrm S^2$ sitting inside
$\mathbb{C}\mathrm P^2$, via the parameterization $\alpha \mapsto
(x(\alpha), y(\alpha), z(\alpha)) = (\,\alpha^2 -
1,\,2\alpha,\,i\,(\alpha^2 + 1)\,)$, for $\alpha \in \mathbb{C}
\cup \{ \infty \}$. Evaluating $P$ on the image of that
parameterization yields a single equation $P(\,\alpha^2 -
1,\,2\alpha,\,i\,(\alpha^2 + 1)\,) = 0$ of degree $2\ell$ in the
single variable $\alpha$, which Mathematica solves quickly and
accurately for the $2\ell$ values of $\alpha$ (counting
multiplicities).  Applying the parameterization gives $2\ell$
solutions $(\hat x_i,\, \hat y_i,\, \hat z_i) = (\,\alpha_i^2 -
1,\,2\alpha_i,\,i\,(\alpha_i^2 + 1)\,)$.  Because the original
equations are real, whenever a point $(\hat x_i,\, \hat y_i,\,
\hat z_i)$ satisfies them, the complex conjugate $\overline{(\hat
x_i,\, \hat y_i,\, \hat z_i)}$ will satisfy them as well. Thus the
$2\ell$ solutions comprise $\ell$ conjugate pairs $\{(\hat x_i,\,
\hat y_i,\, \hat z_i), \overline{(\hat x_i,\, \hat y_i,\, \hat
z_i)} \}$.  Recalling from above that the real and imaginary parts
of each solution are linearly independent, no nontrivial solution
may equal its complex conjugate (or even a scalar multiple
thereof) so the solution pairs are all non-degenerate. Each
solution pair yields a single multipole vector $\pm\bu_i$ via the
cross product of its real and imaginary parts
(Eqn.~(\ref{CrossProduct})).

Must this procedure find all the multipole vectors $\bu_i$, as
opposed to finding some $\bu_i$ several times each while missing
others entirely? Yes.  Pick an arbitrary but fixed $\bu_i$, and
let $(\hat x, \hat y, \hat z)$ be a common solution of $\bu_i
\cdot (x,y,z) = 0 $ and $x^2 + y^2 + z^2 = 0$.  More concretely,
any scalar multiple of $(\hat x, \hat y, \hat z) = (u_y^2 + u_z^2,
- u_x u_y \pm i\,u_z, - u_x u_z \mp i\,u_y)$ is such a solution,
where $(u_x, u_y, u_z)$ are the components of the given $\bu_i$. A
glance at (\ref{succinct}) shows that $(\hat x, \hat y, \hat z)$
must satisfy $P(x,y,z) = 0$ as well, so our algorithm will find
scalar multiples of $(\hat x, \hat y, \hat z)$ and
$\overline{(\hat x, \hat y, \hat z)}$ among the common solutions
to $P(x,y,z) = 0$ and $x^2 + y^2 + z^2 = 0$. Substituting either
$(\hat x, \hat y, \hat z)$ or $\overline{(\hat x, \hat y, \hat
z)}$ into (\ref{CrossProduct}) then gives the desired multipole
vector $\bu_i$, thus proving that our algorithm does indeed find
all of them.

In the non-generic case that some of the $\bu_i$ occur multiple
times each in (\ref{succinct}), does our algorithm get the
multiplicities right?  The easy way to see that it does is to
slightly perturb the polynomial $P(x,y,z)$ so that the $\ell$
multipole vectors $\bu_i$ become distinct.  For the perturbed
polynomial, exactly one pair of roots $\{(\hat x, \hat y, \hat z),
\overline{(\hat x, \hat y, \hat z)}\}$ leads to each $\bu_i$. If
we gradually unperturb the polynomial, the $\bu_i$ fuse back into
their original groupings, and we see that each $\bu_i$ gets mapped
to by the right number of root pairs.

\section{Implementation}
\label{SectionImplementation}

The algorithm, implemented as a Mathematica notebook, is freely
available at {\tt http://www.geometrygames.org/Maxwell/}. The user
specifies a spherical harmonic as an array $\{a_{\ell 0}, \dots,
a_{\ell \ell}\}$ of coefficients in the usual spherical harmonic
decomposition $\sum_{m = -\ell}^{+\ell} a_{\ell m}Y_\ell^m$. Note
that the $a_{\ell m}$ are given explicitly only for $m \geq 0$;
the values for $m < 0$ the defined implicitly via $a_{\ell,-m} =
(-1)^m a_{\ell,+m}^*$.

As an example, the $a_{\ell m}$ for the quadrupole component of
Eriksen~et~al.'s Lagrange Internal Linear Combination (LILC)
cleansing of the first-year WMAP data~\cite{Eriksen2} are
\begin{equation}
  a_{2,m}^{\rm LILC} = \{
    16.30,
    -2.57 + 4.98 i,
    -18.80 - 20.02 i\}.
\end{equation}
Our Mathematica function {\tt AlmToPolynomial} converts the
$a_{2,m}^{\rm LILC}$ to a polynomial
\begin{eqnarray}
  P(x,y,z) =
    -19.67\, x^2 +
    9.39\, y^2 +
    10.28\, z^2 \nonumber\\
    +
    30.94\, x y +
    7.71\, y z +
    3.98\, z x.
  \label{LILC2Polynomial}
\end{eqnarray}
The main algorithm, described in Section~\ref{SubsectionAlgorithm}
and implemented in the Mathematica function {\tt
PolynomialToMultipoleVectors}, yields the Maxwell multipole
vectors
\begin{eqnarray}
  \bu_1 &=& \{\phantom{-}0.975, \phantom{-}0.043, \phantom{-}0.220\}\nonumber\\
  \bu_2 &=& \{\phantom{-}0.634, -0.737, -0.234\}
  \label{LILC2Vectors}
\end{eqnarray}
As an error check, one may feed the multipole
vectors~(\ref{LILC2Vectors}) to the function {\tt
Multipole\-Vectors\-To\-Polynomial} to recover the harmonic
polynomial~(\ref{LILC2Polynomial}).

\subsection{Speed}

The algorithm of Section~\ref{SubsectionAlgorithm}, as implemented
in the function {\tt Polynomial\-To\-Multipole\-Vectors}, runs
exceedingly fast.  On a 2.4~GHz personal computer, the algorithm
computes the multipole vectors $\bu_i$ for randomly generated
$a_{\ell m}$ almost instantaneously for very small $\ell$, and
within a fraction of a second for relevant larger $\ell$
(Table~\ref{Runtime}, middle column).

\begin{table}
  \centering
  \begin{tabular}{|r|c|c|}
  \hline
         &    run time     &   run time   \\
  $\ell$ & without caching & with caching \\
  \hline
    8 &  0.03 s & 0.02 s \\
   16 &  0.2  s & 0.03 s \\
   32 &  0.6  s & 0.1  s \\
   64 &  3    s & 0.4  s \\
  128 & 20    s & 2    s \\
  \hline
\end{tabular}

  \caption{The implementation of the {\tt Polynomial\-To\-Multipole\-Vectors}
  algorithm runs quickly for low values of~$\ell$.  As $\ell$ increases,
  the call to the built-in Mathematica function {\tt Expand}
  for simplifying a polynomial becomes a bottleneck (middle column).
  For Monte Carlo simulations, which require many computations
  for the same~$\ell$, pre-simplifying the spherical harmonics
  and caching them greatly improves performance (righthand column).}
  \label{Runtime}
\end{table}

The algorithm's bottleneck lies in the built-in Mathematica
function {\tt Expand}, with which {\tt
Polynomial\-To\-Multipole\-Vectors} reduces the polynomial in
$\alpha$ (discussed in Section~\ref{SubsectionAlgorithm}) to its
canonical form $c_0 + c_1 \alpha + c_2 \alpha^2 + \dots +
c_{2\ell} \alpha^{2\ell}$ before solving for the roots.  In
particular, simplifying the polynomial to standard form is far
slower than extracting the roots.  When using Monte Carlo methods
to compare the observed WMAP data to large numbers of simulated
CMB skies, one may speed up the multipole computation by
pre-computing the spherical harmonics, making the substitutions
$\{x \mapsto \alpha^2 - 1,\; y \mapsto 2\alpha,\; z \mapsto
i\,(\alpha^2 + 1\,)\}$ directly into the spherical harmonics,
simplifying the polynomials to the canonical form $b_0 + b_1
\alpha + b_2 \alpha^2 + \dots + b_{2\ell} \alpha^{2\ell}$, and
caching the result.  With pre-cached spherical harmonics, the
runtime drops to a very manageable $O(\ell^2)$
(Table~\ref{Runtime}, righthand column).

\subsection{Precision}

\subsubsection{In theory}

The multipole algorithm is quite stable.  To test it,
\begin{enumerate}
  \item begin with a random set of multipole vectors
        $\{\bu_1,\dots,\bu_\ell\}$,
  \item convert the multipole vectors $\{\bu_1,\dots,\bu_\ell\}$
        to a homogeneous polynomial $P$, and then
  \item use $P$ to recompute the multipole vectors
        $\{\bu'_1,\dots,\bu'_\ell\}$.
\end{enumerate}
Table~\ref{NumericalStability} shows that the re-computed
multipole vectors $\{\bu'_1,\dots,\bu'_\ell\}$ agree with the
original multipole vectors $\{\bu_1,\dots,\bu_\ell\}$ to high
precision, even for fairly large values of $\ell$.

\begin{table}
  \centering
  \begin{tabular}{|c|ccccccc|}
  \hline
  $       \ell$       &      2     &      4     &      8     &     16     &     32     &     64     &    128    \\
  \hline
  $\max\angle\bu\bu'$ & $10^{-16}$ & $10^{-16}$ & $10^{-15}$ & $10^{-15}$ & $10^{-14}$ & $10^{-11}$ & $10^{-6}$ \\
   \hline
\end{tabular}
  \caption{The algorithm for computing the $\bu_i$ is robust.
  For each given value of $\ell$, the table shows
  the maximum angle (in radians) between a multipole vector $\bu_i$ and
  the re-computed $\bu_i'$.}
  \label{NumericalStability}
\end{table}

\subsubsection{In practice}

\begin{table}
  \centering
  \begin{tabular}{|c|ccccccccc|}
  \hline
  $\ell$ &  2   &  3   &  4   &  5   &  6   &  7   &  8   &  9   & 10   \\
  \hline
  $\max\angle\bu_{_{\rm DQT}}\bu_{_{\rm LILC}}$ & $16^\circ$ & $3^\circ$ & $8^\circ$ & $6^\circ$ & $13^\circ$ & $16^\circ$ & $10^\circ$ & $12^\circ$ & $8^\circ$ \\
   \hline
\end{tabular}
  \caption{The method used to clean the WMAP data has a significant
  effect on the multipole vectors.  For each value of~$\ell$, the table
  gives the maximum angle (in degrees) between a multipole vector
  computed with the DQT data and the corresponding multipole vector
  computed with the LILC data.}
  \label{CosmologicalStability}
\end{table}

Given the inherent numerical stability of the multipole algorithm,
the real question becomes, how stable are the multipole vectors in
a cosmological context? To answer that question, compute the
low-$\ell$ CMB multipole vectors for the first-year WMAP data,
first using the DQ-corrected Tegmark (DQT) cleaning~\cite{Tegmark}
and then using the Lagrange Internal Linear Combination (LILC)
cleaning~\cite{Eriksen2} of the same raw data. The resulting
multipole vectors differ by about $10^\circ$ from one cleaning to
the other (Table~\ref{CosmologicalStability}).  Thus we conclude
that the computed multipole vectors have physical meaning, but
only roughly.

The most striking entry in Table~\ref{CosmologicalStability} is
the $16^\circ$ maximum difference between the DQT and LILC
quadrupole vectors.  It's tempting to assume that this difference
results from the Doppler correction included in the DQT data but
omitted from the LILC.  However, the maximum difference between
the non-Doppler-corrected Tegmark quadrupole vectors and the LILC
quadrupole vectors is a whopping $37^\circ$.  We must therefore
concede that we know the true quadrupole vectors only very
coarsely.

\section{Applications}
\label{SectionApplications}

\subsection{Quadrupole-octopole alignment}

Schwarz, Starkman, Huterer and Copi offer two variations on their
approach to measuring the quadrupole-octopole
alignment~\cite{Schwarz}. Both variations begin with the two
multipole vectors $\{\bu_{2,1}, \bu_{2,2}\}$ for the quadrupole
and the three multipole vectors $\{\bu_{3,1}, \bu_{3,2},
\bu_{3,3}\}$ for the octopole.

According to the {\it unnormalized version}, a cross product
$\bw_2 = \bu_{2,1} \times \bu_{2,2}$ defines a normal vector to
the {\it quadrupole plane}. Similarly
\begin{eqnarray}
  \bw_{3,1} = \bu_{3,2} \times \bu_{3,3}\nonumber\\
  \bw_{3,2} = \bu_{3,3} \times \bu_{3,1}\nonumber\\
  \bw_{3,3} = \bu_{3,1} \times \bu_{3,2}
\end{eqnarray}
define normal vectors to the three {\it octopole planes}.  The
three dot products $A_i = | \bw_2 \cdot \bw_{3,i} |$, for $i =
1,2,3$, then measure the alignment of the quadrupole plane with
each of the three octopole planes.  Katz and Weeks take the sum $S
= A_1 + A_2 + A_3$, which they find to be unusually high at the
$99.9\%$ level when compared to $10^5$ Monte Carlo
simulations~\cite{KatzWeeks}.

The {\it normalized version} is identical to the unnormalized one,
except that it replaces the previous cross products with {\it unit
length} normal vectors $\bw_2' = |\bw_2| = | \bu_{2,1} \times
\bu_{2,2} |$ and $\bw_{3,i}' = |\bw_{3,i}|$.

So which version should we use?  If our goal is to measure how
well the quadrupole plane aligns with each octopole plane, then
the normalized version works perfectly.  The unnormalized version,
on the other hand, resists easy interpretation.  By retaining the
length of each cross product $\bu_{i,j} \times \bu_{i,j'}$, it
puts more weight on ``well defined planes'' (for which $\bu_{i,j}$
and $\bu_{i,j'}$ are nearly orthogonal and which therefore
robustly determine a normal direction) and less weight on ``poorly
defined planes'' (for which $\bu_{i,j}$ and $\bu_{i,j'}$ are
nearly parallel or anti-parallel and which therefore only weakly
determine a normal direction).  Nevertheless it's not clear what
overall significance the resulting statistic carries, nor whether
this version is a good choice for the quadrupole-octopole
comparison. For the moment let us remain neutral and compute
statistics relative to both versions.

The multipole algorithm of Section~\ref{SubsectionAlgorithm}, as
implemented in the Mathematica code of
Section~\ref{SectionImplementation}, generated a million Monte
Carlo simulations, whose quadrupole-octopole alignments were then
compared to {\it (a)} the DQ-corrected Tegmark (DQT) cleaning of
the first-year WMAP data~\cite{Tegmark} and {\it (b)} the Lagrange
Internal Linear Combination (LILC) cleaning of the first-year WMAP
data~\cite{Eriksen2}.  Table~\ref{QuadrupoleOctopoleResults} shows
the results.

\begin{table}
  \centering
  \begin{tabular}{|r|c|c|}
  \hline
   & unnormalized & normalized \\
   &   version    &  version   \\
  \hline
   DQT  & 99.9\% & 98.7\% \\
  \hline
   LILC & 97.3\% & 98.7\% \\
  \hline
\end{tabular}

  \caption{How unusual is the quadrupole-octopole alignment observed
  in the first-year WMAP data?  For each data set (DQT, LILC) and
  each version of the algorithm (normalizing the lengths of normal
  vectors or not), the table reports what fraction of a million Monte
  Carlo skies had weaker quadrupole-octopole correlations.}
  \label{QuadrupoleOctopoleResults}
\end{table}

The normalized alignment statistic
(Table~\ref{QuadrupoleOctopoleResults}, righthand column) appears
far more robust than the unnormalized alignment statistic
(Table~\ref{QuadrupoleOctopoleResults}, middle column) relative to
small changes in the data, namely switching from the DQT cleaning
of the first-year WMAP data to the LILC cleaning of the same data.
Conceivably dumb luck alone might have given the DQT and LILC data
similar values of the normalized alignment statistic. To test that
possibility, one may compute the same statistics for other data.
Comparing the quadrupole ($\ell=2$) to the hexadecapole ($\ell=4$)
for the same data sets (Table~\ref{QuadrupoleHexadecapoleResults})
again finds the normalized statistic more stable than the
unnormalized one.  Returning to our main data
(Table~\ref{QuadrupoleOctopoleResults}) this suggests that the
normalized statistic measures the quadrupole-octopole alignment
more stably than the unnormalized statistic does.

\begin{table}
  \centering
  \begin{tabular}{|r|c|c|}
  \hline
   & unnormalized & normalized \\
   &   version    &  version   \\
  \hline
   DQT  & 34.6\% & 19.5\% \\
  \hline
   LILC & 17.9\% & 21.0\% \\
  \hline
\end{tabular}
  \caption{Comparing the quadrupole ($\ell=2$)
  to the hexadecapole ($\ell=4$) finds the normalized
  statistic to be more stable than the unnormalized statistic,
  just as in the quadrupole-octopole comparison
  of Table~\ref{QuadrupoleOctopoleResults}.}
  \label{QuadrupoleHexadecapoleResults}
\end{table}

The reason the unnormalized statistic rates the DQT data so much
higher than the LILC data (Table~\ref{QuadrupoleOctopoleResults},
middle column) has little to do with quadrupole-octopole
alignment, but instead reflects the orthogonality of the two
quadru\-pole vectors $\{\bu_{2,1}, \bu_{2,2}\}$.   Specifically,
their cross product $\bw_2 = \bu_{2,1} \times \bu_{2,2}$ has
length $| \bw_2 | = 0.990$ for the DQT data but only $| \bw_2 | =
0.845$ for the LILC data. So even if the LILC octopole bears {\it
exactly} the same relation to the LILC quadrupole plane that the
DQT octopole bears to the DQT quadrupole plane, the sum $S = A_1 +
A_2 + A_3$ will come out $17\%$ higher for the DQT data than for
the LILC data, because
\begin{equation}
  \frac{\;| \bw_2^{^{\rm DQT}} |\;}{\;| \bw_2^{^{\rm LILC}} |\;}
  = \frac{0.990}{0.845} = 1.17.
\end{equation}
Indeed, for the actual data the ratio of the measured sums is
\begin{equation}
  \frac{\;S_{_{\rm DQT}}\;}{\;S_{_{\rm LILC}}\;}
  = \frac{2.395}{2.048} = 1.17.
\end{equation}
Thus the near orthogonality of the quadrupole vectors
$\{\bu_{2,1}, \bu_{2,2}\}$ in the DQT data fully explains the
difference between the two entries in the middle column of
Table~\ref{QuadrupoleOctopoleResults}.  The relative quality of
the quadrupole-octopole alignments plays no role there.

One concludes that the $99.9\%$ value in
Table~\ref{QuadrupoleOctopoleResults} measures the combined
unlikelihood of the quadrupole-octopole alignment {\it and} the
near orthogonality of the DQT quadrupole vectors.  To measure the
quadrupole-octopole alignment alone, one should use the normalized
statistic (Table~\ref{QuadrupoleOctopoleResults}, righthand
column), which for the first-year WMAP data finds the alignment to
be unusual at the $98.7\%$ level.  While this contradicts one
recent claim~\cite{Schwarz,KatzWeeks}, it agrees well with an
earlier estimate (using other methods) that the
quadrupole-octopole alignment is unusual at the 1-in-60
level~\cite{deOliveiraCosta}.

\subsection{Ecliptic plane alignment}

Schwarz~et~al. point out the alignment of the quadrupole and
octopole vectors with the ecliptic plane.  Here we test the
statistical significance of that alignment and obtain results
inconsistent with the $99.8\%$ confidence level claimed in the
preprint version of Ref.~\cite{Schwarz} but consistent with the
revised $\sim 99\%$ confidence level appearing in the final
published version of Ref.~\cite{Schwarz} if one considers only a
one-tailed distribution.

First consider the normal to the quadrupole plane, which in
galactic coordinates is $(-106^\circ,57^\circ)$ for the DQT data
or $(-112^\circ,62^\circ)$ for the LILC data.  Taking the dot
product with the ecliptic pole $\pm(96.4^\circ,29.8^\circ)$ gives
$0.027$ or $0.080$, nominally implying a confidence level of
$97.3\%$ or $92.0\%$, depending on the data set
(Endnote~\footnote{Using the dot product to infer a confidence
level is surprisingly straightforward.  Starting from Archimedes'
observation that axial projection of a cylinder onto an inscribed
sphere preserves areas, it follows, for example, that precisely
$5\%$ of the vectors on the unit sphere yield a magnitude of
$0.95$ or greater when dotted with a fixed unit vector.}).
However, assuming we would have been equally pleased with a dot
product near 1 (meaning the quadrupole plane aligned with the
galactic plane), a two-tailed distribution is appropriate,
dropping the confidence level to $94.6\%$ or $84.0\%$.

Applying the same method to the three octopole planes gives dot
products of $0.523$, $0.045$ and $0.179$ (for the DQT data, in
agreement with Ref.~\cite{Schwarz}) or $0.555$, $0.030$ and
$0.146$ (for the LILC data).  At this point the
Preprint~\cite{Schwarz} computes a raw score which cannot be
interpreted as a confidence level;  to avoid that trap we take a
different approach and examine the sum of the three dot products,
namely $0.747$ (DQT) or $0.731$ (LILC).  Monte Carlo simulation of
$10^5$ Gaussian random skies finds the sum to be larger than that
$95.7\%$ of the time (DQT) or $96.1\%$ of the time (LILC).

Finally, for more direct comparison with the results of
Ref.~\cite{Schwarz}, consider the joint sum of the single
quadrupole dot product along with the three octopole dot products.
Monte Carlo simulation of $10^5$ Gaussian random skies finds this
combined sum to be larger than the observed sum $99.0\%$ of the
time (DQT) or $98.9\%$ of the time (LILC).  This $99\%$ result is
weaker than the erroneous $99.8\%$ claim of
Preprint~\cite{Schwarz} but consistent with the $99.1\%$ value in
the final published Ref.~\cite{Schwarz}.  Allowing for a
two-tailed distribution --- because quadrupole and octopole planes
parallel to the ecliptic would have interested us just as much as
ones perpendicular to the ecliptic --- drops the confidence level
to $98\%$, but even so the result remains valid at better than the
$2\sigma$ level.

\subsection{Planarity of modes}

De~Oliveira-Costa et al. found the WMAP octopole to be unusually
planar at the 1-in-20 level using their
$t$-statistic~\cite{deOliveiraCosta}.  Multipole vectors offer an
independent measure of planarity:  the {\it triple product}
$|\det(\bu_1, \bu_2, \bu_3)|$ of the three octopole vectors
measures the volume of the parallelepiped they span.  When the
octopole vectors are mutually orthogonal, their triple product is
one.  At the other extreme, when the octopole vectors are coplanar
their triple product is zero.  Unfortunately the first-year WMAP
octopole's triple product is only slightly low:   $28\%$ (DQT) or
$30\%$ (LILC) of Gaussian random skies achieve a lower value.
Thus we conclude that De~Oliveira-Costa et al.'s $t$-statistic is
more sensitive to the observed anomaly.

Extending this approach to higher $\ell$, by summing the triple
product of all three-element subsets of the multipole vectors
$\{\bu_1, \dots, \bu_\ell\}$, yields similarly disappointing
results (Table~\ref{Planarity}).  At $\ell = 5$ the triple product
method provides a weak $2\sigma$ confirmation of Eriksen et. al's
$3\sigma$ observation of non-planarity~\cite{Eriksen2}, but
otherwise the data in Table~\ref{Planarity} seem consistent with a
flat distribution on the interval $[0\%,100\%]$. Again we conclude
that the $t$-statistic better detects planarity or lack thereof.

\begin{table}
  \centering
  \begin{tabular}{|c|cccccccc|}
  \hline
  $\ell$ &  3   &  4   &  5   &  6   &  7   &  8   &  9   & 10   \\
  \hline
  DQT    & 28\% & 80\% & 89\% & 24\% & 25\% & 58\% & 82\% & 69\% \\
  LILC   & 30\% & 58\% & 97\% & 33\% & 12\% & 72\% & 48\% & 34\% \\
  \hline
\end{tabular}
  \caption{For each data set (DQT, LILC) and each value of $\ell$
  (3..10), the table shows what percentage of $10^5$ Monte Carlo
  simulations yielded a more planar set of multipole vectors.}
  \label{Planarity}
\end{table}

\section{Conclusions}

Maxwell's multipole vector construction sheds much light on
spherical harmonics and offers a clean and simple approach to
quantifying alignments among the low-$\ell$ CMB modes.  In the
case of the quadrupole-octopole alignment, simulations
convincingly show the alignment to be unusual at the $98.7\%$
level, corroborating earlier estimates of a 1-in-60
alignment~\cite{deOliveiraCosta} while showing recent reports of
1-in-1000~\cite{Schwarz,KatzWeeks} to have depended on the near
orthogonality of the quadrupole vectors in the DQT data. The
alignment of the quadrupole and octopole vectors with the ecliptic
plane is confirmed at better than the $2\sigma$ level. \\

\section*{Acknowledgments}

I thank Mark Dennis whose paper~\cite{Dennis1} brought Maxwell's
work to my attention and whose subsequent preprint~\cite{Dennis2}
offers additional insights, and I thank Ben Lotto and John
McCleary for helpful conversations.

\end{document}